# GAMIFICATION AS A DATA ACQUISITION STRATEGY FOR AI TRAINING IN NEUROGAMES USING ACTIVE BCI - A VIDEOGAME INDUSTRY PERSPECTIVE

**ABSTRACT:**
The nascent field of neurogames relies on active Brain-Computer Interface input to drive its game mechanics. Consequently, users expect their conscious will to be meaningfully reflected on the virtual environment they're engaging in. Additionally, the videogame industry considers it paramount to provide gamers with seamless experiences to avoid disrupting their state of flow. Thus, this paper suggests gamification as a strategy to camouflage the often fatiguing data acquisition process in Machine Learning from neurodata so that neurogamers can further immerse themselves in the virtual experience while Artificial Intelligence models benefit from data taken in reproducible contexts.

**INTRODUCTION:**
While observation of certain emotional states can be achieved by neuroadaptively correlating users' brainwave bands [1], it is imperative to collect data of users' brain activity in order to correlate them to specific commands for use as game mechanics in a ludic environment where an active Brain Computer-Interface (BCI) reflects users' conscious will through meaningful multimedial feedback. Evidence suggests this to be a laborious process [2] that, when rushed or incomplete, may lead to a faulty Machine Learning (ML) model that results in unsatisfactory neurofeedback, break in immersion and precipitated rejection of the neurogaming experience [3]. Indeed, the more complex the initial setup of the neurogame, the less likely it is to retain players, nowadays so accustomed to low barriers of entry for mainstream games. In fact, scientific review points out that gamification of neurodata acquisition yields better results than a dry, clinical approach [4]. Consequently, minimizing, automating and camouflaging the setup of neurogaming systems should be taken for granted as a best practice. Under such a design philosophy informed by scientific review, neurogame developers should endeavor to recontextualize the neurodata acquisition process in an unobtrusive format through gamification and/or actual game mechanics. Thus, this paper proposes gamifying the neurodata acquisition process by using two standard tools in the videogame industry, namely: tutorials and event-driven programing.

**STRATEGIES:**
The following strategies are meant to be applied to any gaming platform which can indirectly or, preferably, directly interface with neurotechnology through a game engine which, ideally, would be in charge of data acquisition, data processing and Artificial Intelligence (AI) training. They can be adapted to less centralized systems, so long as the player is not in charge of setting them up. Research suggests that tutorials should be integrated with gameplay as part of the overall ludonarrative [5], which offers a reproducible context to integrate initial gathering of data. A consonant ludonarrative (game story and mechanics working in tandem) [6] may use the game's narrative to integrate the need for calibration and data acquisition as players' are instructed in the use of the game mechanics dependent on active BCI. The game may further reinforce proper BCI usage as part of the narrative, as artifacts and lowered performance have a direct ludonarrative impact on the game. Thus, players could become more immersed even when the BCI underperforms. Until BCIs can technically blur the lines between reality and virtuality, game developers must bridge this gap with a compelling narrative. As an example, consider the ludonarratives (story and mechanics reinforcing each other) present in the following synopsis:

> A psychic student uses a headset to enhance their psychic skills. If the headset's electrodes do not have proper contact, the psychic effects are diminished. The more the student uses

*Manuscript submitted and approved for NAT'25, Berlin.*
*This manuscript was withdrawn because lack of funding prevented the author from taking part in the conference.*

this headband, the more attuned it becomes to their psychic intentions. The first psychic exercise is to make a boulder explode while standing close to it. On the first few tries, boulders merely crumble laboriously. As gameplay unfolds, affected boulders explode with increased violence. Other psychic skills are obtained and trained in a similar progression.

This example also showcases the application of event-driven programming during gameplay. In order to implement the aforementioned ludonarratives, it is necessary to use events to trigger, at the same time, the game mechanic (e.g. boulder destruction) and the recording of neurodata to add to the database for further ML training. This approach benefits from the inclusion of audiovisual cues to elicit memories whose brainwave signature could become part of the ML model. These events can be triggered in different manners, according to the tools available in each game engine. For instance, a collider can trigger an event signaling the presence of a player in an area of interest. In the ludonarrative example above, the game would start comparing the player's brainwave pattern with the trained database to predict the possibility that the player intends to destroy a boulder. A raycasting algorithm could track how far the player's avatar is from the boulder in order to increase or multiply the aforementioned prediction calculated by the AI. Or, to give more agency to the player, the game controller or keyboard could map a button to trigger the use of the boulder destruction game mechanic. While the button is pressed, the game would make use of the predictions being calculated by the AI.

**DISCUSSION:**

Although it might be tempting to try to completely replace the manual game controls with a BCI, it is important to remember that manual game controls have been developing and perfecting since 1958 [7], whereas BCI-driven controls for gaming have not yet had enough time to perfect their ergonomics and responsiveness to a degree where it may be preferable to use your thoughts before your hands as a game mechanic [8]. Thus, extending the controller instead of replacing it might be the path of least resistance, as exemplified previously. That being said, an enterprising developer who would place a player in a game driven exclusively by neuro- and biofeedback, may need to offer a compelling narrative which encourages the player to persevere despite a considerable learning curve—for both the player and the AI; especially if the target audience is not comprised exclusively of people with severely reduced mobility who also happen to have experience with the fatiguing endeavour of mapping their neural activity to BCI-driven aids [9].

Whatever the approach to neurogames' game mechanics, it is important to develop them with a working understanding of the capabilities and limitations of the BCI in use, as well as a proper understanding of game design theory and practice. If the approach is centered around neurotechnology but ignores what makes a game engaging to a wider audience, it risks commercial failure. An approach that designs game mechanics that ignore the way neurotechnology works, risks delaying the development process or altogether failing to deliver. These risks could be mitigated by designing and testing the tutorial and its underlying event-driven mechanics at an early stage, solidifying engaging mechanics and expunging detrimental design biases. A multi-disciplinary approach to neurogame design and development may also decrease these risks.

**CONCLUSION:**

In summary, neurogames that use ML and AI to process brainwaves as part of their game mechanics would benefit from the gamification of the data acquisition process, so as to not only make them more accessible to end users', but also to motivate them to sustainably engage in an otherwise laborious process. Implementing consonant ludonarratives throughout the game reinforced by tutorials and event-driven programming might prove to be advantageous strategies for mainstream neurogame design. Scientific research may also benefit from these strategies, especially in a context where the subject onboarding process and the ML processes are automated by a game engine, thus streamlining researchers' methodologies. Proper implementation of these strategies requires a multi-disciplinary approach, whether it means including neuroscientists in game development studios or including game design interns in research departments.

*Manuscript submitted and approved for NAT'25, Berlin.*
*This manuscript was withdrawn because lack of funding prevented the author from taking part in the conference.*

*Manuscript submitted and approved for NAT'25, Berlin.*
*This manuscript was withdrawn because lack of funding prevented the author from taking part in the conference.*